# Analysis of Nonlinear Electromagnetic Metamaterials

Ekaterina Poutrina, Da Huang, David R. Smith

*Center for Metamaterials and Integrated Plasmonics and Department of Electrical and Computer Engineering, Duke University, Box 90291, Durham, NC 27708*

**Abstract**

We analyze the properties of a nonlinear metamaterial formed by integrating nonlinear components or materials into the capacitive regions of metamaterial elements. A straightforward homogenization procedure leads to general expressions for the nonlinear susceptibilities of the composite metamaterial medium. The expressions are convenient, as they enable inhomogeneous system of scattering elements to be described as a continuous medium using the standard notation of nonlinear optics. We illustrate the validity and accuracy of our theoretical framework by performing measurements on a fabricated metamaterial sample composed of an array of split ring resonators (SRRs) with packaged varactors embedded in the capacitive gaps in a manner similar to that of Wang et al. [*Opt. Express* **16**, 16058 (2008)]. Because the SRRs exhibit a predominant magnetic response to electromagnetic fields, the varactor-loaded SRR composite can be described as a magnetic material with nonlinear terms in its effective magnetic susceptibility. Treating the composite as a nonlinear effective medium, we can quantitatively assess the performance of the medium to enhance and facilitate nonlinear processes, including second harmonic generation, three- and four-wave mixing, self-focusing and other well-known nonlinear phenomena. We illustrate the accuracy of our approach by predicting the intensity-dependent resonance frequency shift in the effective permeability of the varactor-loaded SRR medium and comparing with experimental measurements.

## 1. Introduction

Metamaterials consist of arrays of magnetically or electrically polarizable elements. In many common configurations, the functional metamaterial elements are planar, metal patterns, on which currents are induced that flow in response to incident electromagnetic fields. Since these effective circuits are typically much smaller than the wavelengths over which the metamaterial is expected to operate, the polarizability of each element can be determined by applying a relatively straightforward quasi-static circuit model in which standard circuit parameters such as inductance, capacitance and resistance are introduced [1-4]. From the circuit point-of-view, it is then easily appreciated that the much larger range of properties routinely observed in metamaterials—including negative permittivity, permeability and refractive index—relates directly to the underlying electrical "LC" resonances that the metamaterial circuits support. The strong, resonant response of the metamaterial circuit when coupled to applied fields translates to a strong polarizability in the language of material science. Applying homogenization methods to a collection of metamaterial elements, we thus obtain conceptually a continuous medium whose effective constitutive parameters can be controlled by the design of the metamaterial geometry. With the expanded palette of response available in metamaterials, electromagnetic structures and devices can be designed and demonstrated, negative index materials [5-7] and transformation optical media [8, 9] being striking examples.

Though circuit methods are usually associated with radio-frequency and microwave applications, the circuit description for metamaterials provides an accurate model for higher frequency and optical metamaterials as well. At frequencies significantly lower than the plasma frequency of a conductor, it is possible to associate inductance with a conducting path and capacitance with gaps in the conducting path. However, the mechanism of inductance changes towards visible wavelengths, since the inductance becomes dominated by the inertial response of the charge carriers [10].

A distinct property of metamaterials is that the capacitive regions strongly confine and enhance the local electric field. The capacitive region of a metamaterial element thus provides a natural entry point at which exogenous frequency-dispersive, active, tunable or nonlinear materials can be introduced, providing a mechanism for coupling the geometric LC resonance to other fundamental material properties [1]. The hybridization of metamaterials with semiconductors [11], ferromagnetic materials [12] and other externally tunable materials [13] has already expanded the realm of metamaterials from the passive, linear media initially considered.

Nonlinear materials represent a potentially useful class of materials to consider for integration into metamaterial composites. Many crystals and polymers exhibit substantial nonlinearities that are exploited across the spectrum for devices [14]. In conventional materials, the linear and nonlinear susceptibilities and their dispersions are intrinsically set, for example, by the fundamental anharmonic resonances associated with molecular systems. In metamaterial hybrid composites, however, those susceptibilities can potentially be altered independently with considerable control, retaining and even enhancing the nonlinear response of the embedded material while engineering the propagation characteristics of the otherwise linear metamaterial structure. This independent control provides an alternative route to the optimization of nonlinear materials, which form the basis of such optical devices such as mixers, frequency doublers, and modulators [15].

The incorporation of inherently nonlinear materials into the capacitive gaps of split ring resonators was considered theoretically by Zharov et al. [16], who predicted an intensity dependent resonance frequency of the composite medium with associated bistability. The nonlinear phenomena such as harmonic generation [17,18], parametric down-conversion [19], or tunability [20,21] have been demonstrated experimentally either in planar metamaterial structures or in a single unit cell element. Many current studies rely on the theoretical analysis of wave propagation in the structures incorporating a negative index material (NIM), assuming a homogeneous NIM layer with a presumed values of linear and nonlinear response [22- 27]. In this way, a diverse spectrum of nonlinear phenomena has been analyzed and a variety of novel effects has been predicted arising from the specific electromagnetic properties that the negative-index medium possess. Some examples include the predicted enhancement of interface second-harmonic generation near the zero-n gap of a negative-index Bragg grating [23, 24], the enhancement of the parametric processes in binary metamaterials [25], the possibility of satisfying the phase matching condition for the counter-propagating waves for the SH generation in a NIM [26], optical bistability in a nonlinear optical coupler with a negative index channel [27], or gap solitons in a nonlinear negative-index cavity [28].

For practical applications, the design and optimization of nonlinear metamaterial-based devices requires a more quantitative approach, one that relates the particular metamaterial geometry incorporating nonlinear elements to the nonlinear properties of the resulting effective medium. Steps towards this goal have been taken in Refs. [25, 29], where the possibility of representing the nonlinear response of a metamaterial with diode insertions in terms of the effective second-order susceptibility has been discussed. Our goal here is to extend the general analytical framework for a metamaterial based on nonlinear elements integrated into the underlying circuit of each unit cell. Our approach is to expand the effective medium polarizability in a power series in terms of the applied field amplitude, thus relating the response of the metamaterial to the strength of the applied field, in the same manner as is done for nonlinear crystals [14]. To emphasize the analogy between the anisotropic metamaterial hybrid media and natural crystals, we refer to the metamaterial structures described herein as *metacrystals*.

For a system close to resonance, which is the case for a resonant metamaterial, care should be taken regarding convergence of the series at high powers [14]. Regardless of this limitation, the series expansion representation is particularly useful since it allows the interpretation and quantitative prediction of an immense variety of nonlinear phenomena known to originate from certain orders of nonlinear response [14, 30]. As an example, we provide an estimate of the applicable power levels using the varactor- loaded SRR (VLSRR)-based medium. Within the range of applicability the approach allows the development of the model of a homogeneous analog nonlinear medium that aligns with the same concept in nonlinear optics.

In the following section, we introduce the basic circuit model description of a nonlinear metameterial possessing a magnetic resonant response, determine the linear constitutive parameters via an analytic effective medium theory, and derive expressions for the second and the third order magnetic susceptibilities characterizing the nonlinear medium formed by the metacrystal. While there are many materials that could serve as the nonlinear component of a hybrid metamaterial composite, at low frequencies (below a few gigahertz), the low cost and wide availability of varactor diodes makes them a convenient source for integration. The use of varactor diodes as a means of implementing tunability [20] and nonlinear response [31] in metamaterial hybrid media have been demonstrated experimentally. We relate our theory to a particular case of VLSRR-based medium in Section 3. In Section 4, we discuss the range of the applicability and the accuracy of the solution. To illustrate the accuracy of our analytical expressions for the nonlinear susceptibilities, we present the results of experimental measurements on a VLSRR structure and compare the observed nonlinear response to that predicted in Section 5. Both the theoretical and the experimental results are summarized in Section 6.

## 2. Theoretical description
### 2.1. *General expressions for the nonlinear metacrystal*

We consider here a nonlinear metacrystal formed from resonant circuit elements that couple strongly to the magnetic field. Although we specifically consider the split ring resonator (SRR) medium [1-3], the general circuit model here should apply to other effective magnetic metacrystals such as that formed from cut wire pairs [32,33] or the fishnet structure [34, 35]. The principal geometry of an SRR structure, which is essentially a current loop with a capacitive gap that breaks the otherwise continuous current path, is illustrated in Fig. 1a, with its equivalent inductively driven RLC circuit model shown in Fig. 1b. The induced

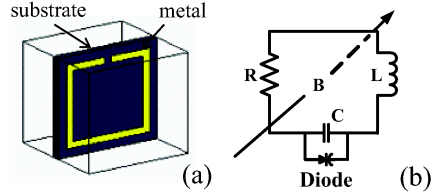

**Fig. 1 (a) Split-ring resonator geometry; (b) equivalent effective circuit model**

electromagnetic force, $\varepsilon(t)$, drives the circuit producing a current $I$ that satisfies

$$L\frac{dI}{dt} + R_s I + V_D = \varepsilon(t) = -\frac{\partial \Phi_m}{\partial t}. \tag{1}$$

In Eq. (1), $L$, $R$, and $V_D = \frac{1}{C}\int_0^t I(\tau)d\tau$ are, respectively, the distributed inductance, distributed resistance, and the induced voltage across the effective capacitor $C$ of the circuit, with the particular interpretation of these parameters determined by the geometry and the frequency range considered. The linear circuit parameters can be estimated from analytical approximations; however, as will be seen later in this section, the parameters can also be obtained by curve fitting the numerically simulated effective permeability. The latter method, based on well-established numerical retrieval procedures, has proven extremely accurate for linear materials and avoids any reliance on analytical approximations [36].

Generally, any or all of the effective circuit parameters ($R$, $L$, or $C$) can possess a nonlinear response to the amplitude of the induced electromagnetic force (and, hence, to the power of the applied field), becoming functions of the amplitude of charge oscillations in the circuit [37]. Introducing the time-dependent charge $Q(t) = \int_0^t I(\tau)d\tau$, Eq. (1) can be written in terms of the normalized charge $q \equiv Q/C_0$, $C_0$ being the value of the capacitance in the linear regime (at low powers), as

$$\frac{d^2 q}{dt^2} + \gamma\frac{dq}{dt} + \omega_0^2 F(q,\dot{q}) = \omega_0^2 \varepsilon(t), \tag{2}$$

where $\omega_0^2 \equiv 1/L_0 C_0$ denotes the resonant frequency of the circuit in the linear regime, $\gamma \equiv R_0/L_0$ the linear damping constant, and $R_0$ and $L_0$ the linear values of the resistance and the inductance. Equation 2 is that of a driven oscillator, with the term $F(q,\dot{q})$ representing a normalized nonlinear restoring force. We note that, while the above formal expression for $\gamma$ only accounts for the resistive loss in the effective circuit, in practice, all loss sources, including the losses in the substrate will contribute to the value of $\gamma$. This value can be estimated numerically, for example, by the decay time of the free oscillations of the initial charge at the effective capacitor [38].

Although nonlinear response could be introduced into any of the circuit parameters, the field enhancement within the capacitive gaps suggests that the nonlinearities of materials or components will be best leveraged by integrating them into the high-field regions of the circuit. We thus assume that only the capacitance $C(q)$ has a nonlinear response, so that Eq. 2 then reduces to [21]

$$\frac{d^2 q}{dt^2} + \gamma \frac{dq}{dt} + \omega_0^2 V_D(q) = \omega_0^2 \varepsilon(t) \tag{3}$$

where $V_d$ is the voltage across the effective capacitor.

We expect that at low power levels the nonlinear response will be generally weaker than the linear response, so that we are justified in expanding the voltage in a Taylor series, retaining a limited number of terms in $q$. In this section, we write the terms of the expansion to third order, with higher (up to fifth order) terms considered in the Appendix in connection with discussing the limits of convergence. The expansion can be written as

$$V_D(q) = q + aq^2 + bq^3, \tag{4}$$

and Eq. (3) then takes the standard form of a nonlinear oscillator:

$$\ddot{q} + \gamma \dot{q} + \omega_0^2 q + \alpha q^2 + \beta q^3 = \omega_0^2 \varepsilon(t), \tag{5}$$

where $\alpha \equiv \omega_0^2 a$ and $\beta \equiv \omega_0^2 b$.

It can be shown [37] that Eq. 2 can equally be brought to the form of Eq. 5 if the effective inductance, or both the inductance and the capacitance were to possess a nonlinear response, with the only difference occurring in the particular expressions for the nonlinear coefficients $a$ and $b$. The form of the equation would also be similar in the case of a nonlinear resistive, with the Taylor expansion in terms of the derivative of the normalized charge.

The nonlinear oscillator form of Eq. 5 can be solved by standard methods. Following Ref. [14], we assume that the response of the oscillator will be in proportion to the strength of the applied field, motivating a perturbative expansion of the form

$$\tilde{q}(t) = \lambda \tilde{q}^{(1)}(t) + \lambda^2 \tilde{q}^{(2)}(t) + \lambda^3 \tilde{q}^{(3)}(t) + \cdots \tag{6}$$

Here, $\lambda$ is a dimensionless perturbation parameter ($\lambda \leq 1$) indicating the field strength $\lambda \varepsilon(t)$. The tilde sign signifies variables that are time-dependent. Substituting Eq. 6 into Eq. 5 and equating terms of the same order in $\lambda$, one can obtain the following system of coupled equations

$$\ddot{\tilde{q}}^{(1)} + \gamma \dot{\tilde{q}}^{(1)} + \omega_0^2 \tilde{q}^{(1)} = \omega_0^2 \tilde{\varepsilon}(t) \tag{7a}$$

$$\ddot{\tilde{q}}^{(2)} + \gamma \dot{\tilde{q}}^{(2)} + \omega_0^2 \tilde{q}^{(2)} + \alpha \left[\tilde{q}^{(1)}\right]^2 = 0 \tag{7b}$$

$$\ddot{\tilde{q}}^{(3)} + \gamma \dot{\tilde{q}}^{(3)} + \omega_0^2 \tilde{q}^{(3)} + 2\alpha \tilde{q}^{(1)} \tilde{q}^{(2)} + \beta \left[\tilde{q}^{(1)}\right]^3 = 0 \tag{7c}$$

Equations 7a, 7b, and 7c describe, respectively, the first (linear), second and third order responses of the normalized charge. Note that the first term in Eq. 7 is the same whether we assume a linear or nonlinear oscillator and that all subsequent higher order terms depend on the amplitude of the lower terms.

Because the SRR exhibits predominantly magnetic response, we assume the incident driving field corresponds to the magnetic component of the linear-polarized incident field, or $\tilde{\mathbf{B}}(t) = \hat{\mathbf{y}} \sum_{n=1}^{\Lambda} B_{r_y}(\omega_n) \cos(kz - \omega_n t) = \hat{\mathbf{y}} \sum_{n=1}^{\Lambda} B_y(\omega_n) e^{-i\omega_n t} + c.c.$, where the subscript $y$ indicates the $y$-component of the magnetic field $\mathbf{B}$, and $B_y(\omega_n) \equiv 0.5 B_{r_y}(\omega_n) e^{ikz}$ with the factor of 1/2 arising from the reality of the field. The axis of the circuit is taken to lie along the $y$-direction. We thus assume that the incident field can be expressed as a sum of $\Lambda$ components, each of which oscillates with an angular frequency $\omega_n$. According to Faraday's law, the driving electromagnetic force $\varepsilon(t)$ is related to the amplitude of the incident magnetic flux according to $\tilde{\varepsilon}(t) = iA \sum_{n=-\Lambda}^{\Lambda} \omega_n B_y(\omega_n) e^{-i\omega_n t}$, where $A$ represents the effective area enclosed by the circuit. In the above expression, the summation is taken over both positive and negative frequencies, where we have assumed $B(-\omega_n) = B^*(\omega_n)$ to arrive at a more compact notation. Henceforward we omit the subscript $y$.

We seek a solution to Eqs. 7 of the form

$$\tilde{q}^{(1)}(t) = \sum_{n=-\Lambda}^{\Lambda} q^{(1)}(\omega_n) e^{-i\omega_n t} \tag{8a}$$

$$\tilde{q}^{(2)}(t) = \sum_r q^{(2)}(\omega_r) e^{-i\omega_r t} \tag{8b}$$

$$\tilde{q}^{(3)}(t) = \sum_s q^{(3)}(\omega_s) e^{-i\omega_s t} \tag{8c}$$

where $\Lambda$ is the number of distinct waves in the applied field, $\omega_r \equiv \omega_n + \omega_m$, $\omega_s \equiv \omega_n + \omega_m + \omega_p$, and the summations are taken over both positive and negative frequencies, with $n$, $m$, and $p$ each taking values between $\pm\Lambda$. Note that, according to this notation, $r$ takes the values of all possible combinations of $m + n$ leading to the combinatorial frequency $\omega_r$; for example, it takes the values of $\pm 2$ and 0 if $\Lambda = 1$; similar considerations hold for the summation over $s$ in Eq. 9c.

Using Eq. 8a in Eq. 7a, we can write the following expression for each $\omega_n$

$$\ddot{q}^{(1)}(\omega_n) + \gamma \dot{q}^{(1)}(\omega_n) + \omega_0^2 q^{(1)}(\omega_n) = i\omega_0^2 A \omega_n B(\omega_n), \tag{9}$$

from which we obtain

$$q^{(1)}(\omega_n) = i \frac{\omega_0^2 \omega_n A B(\omega_n)}{\omega_0^2 - \omega_n^2 - i\gamma\omega_n} \equiv i \frac{\omega_0^2 \omega_n A B(\omega_n)}{D_n}, \tag{10}$$

where we have defined the resonant denominator $D_n \equiv \omega_0^2 - \omega_n^2 - i\gamma\omega_n$. Using Eqs. (8a) and (10), Eq. (7b) can be written as

$$\ddot{\tilde{q}}^{(2)} + \gamma \dot{\tilde{q}}^{(2)} + \omega_0^2 \tilde{q}^{(2)} = \sum_{m,n} \frac{a\omega_0^6 \omega_n \omega_m A^2 B(\omega_n) B(\omega_m)}{D_n D_m} e^{-i(\omega_n + \omega_m)t} \tag{11}$$

where the indices *m* and *n* vary between ±Λ. The right hand side of Eq. 11 contains many combinatorial frequencies because of the summation over *m* and *n*. Assuming the form of the solution given by Eq. 8b, Eq. 11 be transformed into a set of independent equations for each frequency $\omega_r \equiv \omega_n + \omega_m$, with the second-order response $q^{(2)}(\omega_r)$ at each frequency $\omega_r$ satisfying the equation

$$\left(-\omega_r^2 - i\gamma\omega_r + \omega_0^2\right) q^{(3)}(\omega_r) e^{-i\omega_r t} = \sum_{(mn)} \frac{a\omega_0^6 \omega_n \omega_m A^2 B(\omega_n) B(\omega_m)}{D_n D_m} e^{-i(\omega_n+\omega_m)t} \tag{12}$$

In Eq. 12, the brackets under the sum indicate that the sum $\omega_n + \omega_m$ remains fixed while the indices *n* and *m* vary. From Eq. 12, the resulting expression for the second-order response at each $\omega_r$ is

$$q^{(2)}(\omega_r) = a\omega_0^6 A^2 \sum_{(nm)} \frac{\omega_n \omega_m B(\omega_n) B(\omega_m)}{D_n D_m D_{n+m}} \tag{13}$$

The general solution for the second order-response is given by Eq. 8b, summing Eq. 13 over all possible values of $\omega_r$.

Following the same steps as above and using Eq. 10 and 13 in combination with Eqs. 8 in Eq. 7c, we obtain for the third-order response at a frequency $\omega_s \equiv \omega_n + \omega_m + \omega_p$

$$\left(-\omega_s^2 - i\gamma\omega_s + \omega_0^2\right) q^{(3)}(\omega_s) e^{-i\omega_s t} = -i2a^2 \omega_0^{10} A^3 \sum_{(mnp)} \frac{\omega_n \omega_m \omega_p B(\omega_n) B(\omega_m) B(\omega_p)}{D_n D_m D_{n+m} D_p} e^{-i(\omega_n+\omega_m+\omega_p)t}$$

$$+ ib\omega_0^8 A^3 \sum_{(mnp)} \frac{\omega_n \omega_m \omega_p B(\omega_n) B(\omega_m) B(\omega_p)}{D_n D_m D_p} e^{-i(\omega_n+\omega_m+\omega_p)t}, \tag{14}$$

The expression for $q^{(3)}(\omega_s)$ then follows directly from Eq. 14. However, the factor $D_{n+m}$ in the first term on the right hand side of Eq. 14 results from the second order response $q^{(2)}(\omega_r)$ at frequency $\omega_r \equiv \omega_n + \omega_m$ with *n* and *m* taking any values between ±Λ that produce in combination the frequency $\omega_r$. To account for the various contributions, this factor must be permuted over all possible combinations of indices *n* and *m*. We arrive then at the following expression for the third order response at the frequency $\omega_s$ accounting for the second and the third order nonlinearity:

$$q^{(3)}(\omega_s) = ib\omega_0^8 A^3 \sum_{(mnp)} \frac{\omega_n \omega_m \omega_p B(\omega_n) B(\omega_m) B(\omega_p)}{D_n D_m D_p D_s} - i\frac{2}{3} a^2 \omega_0^{10} A^3 \sum_{(mnp)} \frac{\omega_n \omega_m \omega_p B(\omega_n) B(\omega_m) B(\omega_p)}{D_n D_m D_p D_s} \left[\frac{1}{D_{n+m}} + \frac{1}{D_{n+p}} + \frac{1}{D_{m+p}}\right]. \tag{15}$$

The complete solution to Eq. 7c is given by Eq. 8c with $q^{(3)}(\omega_s)$ determined from Eq. 15.

Equations 10, 13, and 15, in combination with Eqs. 8, provide general expressions for the first, second, and third order responses, respectively, accounting for the second and the third order nonlinearities. These expressions can be used to derive the first, second, and third order magnetic susceptibilities that will characterize the metacrystal.

For a dilute medium, the magnetization, $\tilde{M}$, is approximately related to the magnetic moment $\tilde{m}$ according to $\tilde{M}(t) = N\tilde{m}(t)$, where *N* is the volume density of moments. The magnetic dipole moment of the effective circuit encompassing the effective area *A* is given approximately as $\tilde{m}^{(1)}(t) = I^{(1)} A = \dfrac{d\tilde{Q}(t)}{dt} A = A C_0 \dfrac{d\tilde{q}(t)}{dt}$. Following the perturbative expansion for $q(t)$ given by Eqs. 6 and 8, the magnetization can be written as

$$\tilde{M}_y(t) = -iNAC_0 \left[ \sum_{n=-\Lambda}^{\Lambda} \omega_n q^{(1)}(\omega_n) e^{-i\omega_n t} + \sum_r \omega_r q^{(2)}(\omega_r) e^{-i\omega_r t} + \sum_s \omega_s q^{(3)}(\omega_s) e^{-i\omega_s t} + \ldots \right] \quad (16)$$

where, as before, $\omega_r \equiv \omega_n + \omega_m$, $\omega_s \equiv \omega_n + \omega_m + \omega_p$, and each of indices $n$, $m$, and $p$ take values between $\pm\Lambda$.

We assume that the magnetization can be expressed as a power series in terms of the strength of the applied field, according to:

$$\tilde{M}_y(t) = \sum_{n=-N}^{N} M_y^{(1)}(\omega_n) e^{-i\omega_n t} + \sum_r M_y^{(2)}(\omega_r) e^{-i\omega_r t} + \sum_s M_y^{(3)}(\omega_s) e^{-i\omega_s t} + \ldots \quad (17)$$

where

$$M_y^{(1)}(\omega_n) = \chi_{yy}^{(1)}(\omega_n) H_y(\omega_n), \quad (18a)$$

$$M_y^{(2)}(\omega_r) = \sum_{(mn)} \chi_{yyy}^{(2)}(\omega_r; \omega_m, \omega_n) H_y(\omega_m) H_y(\omega_n) \quad (18b)$$

$$M_y^{(3)}(\omega_s) = \sum_{(mnp)} \chi_{yyyy}^{(3)}(\omega_s; \omega_m, \omega_n, \omega_p) H_y(\omega_m) H_y(\omega_n) H_y(\omega_p) \quad (18c)$$

and where we have adopted the conventional notation for the arguments of the nonlinear susceptibility in which the first frequency term is the sum of the subsequent frequency arguments [14]. The subscript $y$ in this notation refects the cartesian coordinates of each participating field component and of the resulting magnetizaton, which are all polarized along the $y$-axis in the present case.

Equating the terms with equal powers of the exponents in Eqs. 16 and 17, and employing Eqs. 10, 13, and 15 for $q^{(1)}(\omega_n)$, $q^{(2)}(\omega_r)$, and $q^{(3)}(\omega_s)$, we obtain, respectively, the linear, second- and third-order magnetic susceptibilities as follows

$$\chi_{yy}^{(1)}(\omega_n) = \dfrac{N\omega_0^2 \omega_n^2 A^2 \mu_0 C_0}{D(\omega_n)}, \quad (19)$$

$$\chi_{yyy}^{(2)}(\omega_r; \omega_m, \omega_n) \doteq -ia \dfrac{\omega_0^6 (\omega_n + \omega_m) \omega_n \omega_m \mu_0^2 A^3 NC_0}{D_n D_m D_{n+m}}, \quad (20)$$

$$\chi_{yyyy}^{(3)}(\omega_r; \omega_n, \omega_m, \omega_p) = -\dfrac{2}{3} a^2 A^3 \dfrac{\omega_0^{10}(\omega_n + \omega_m + \omega_p)\omega_n \omega_m \omega_p \mu_0^3 A^4 NC_0}{D_n D_m D_p D_s}\left[\dfrac{1}{D_{n+m}} + \dfrac{1}{D_{n+p}} + \dfrac{1}{D_{m+p}}\right] + b\dfrac{\omega_0^8 (\omega_n + \omega_m + \omega_p)\omega_n \omega_m \omega_p \mu_0^3 A^4 NC_0}{D_n D_m D_p D_s}, \quad (21)$$

where each frequency can take both positive and negative values with the indices $n$, $m$, and $p$ each varying between $\pm\Lambda$; $\mu_0$ is the permeability of vacuum.

From an inspection of Eq. 19, we see that the linear susceptibility of the SRR can be expressed in terms of its geometrical and electrical parameters as

$$\chi_{yy}^{(1)}(\omega_n) = \dfrac{F\omega_n^2}{D_n} \quad (22)$$

where $F \equiv \omega_0^2 \mu_0 A_0^2 C_0$, in qualitative agreement with more detailed analytical studies [3]. According to Eq. 23, we do not seek more accurate analytical expressions for the effective circuit parameters, since we can easily find the linear properties through numerical retrievals, performing a fitting to determine the unknown coefficients $F$ and $\gamma$ (entering through the factor $D_n$) in Eq. 22. The higher order terms do not add to this parameter set, so in practice it suffices to have the initial expression in Eqs. 19 or 22.

2.2. *Examples for particular combinatorial frequencies.*

As an example of the use of Eqs. 19-21, we write in a more explicit form the susceptibilities for some particular combinatorial frequencies for the second and the third order processes. For example, for the process of second-harmonic generation we have $\omega_m = \omega_n$ and, from Eq. 20:

$$\chi^{(2)}_{yyy}(2\omega_n; \omega_n, \omega_n) = -ia \frac{2\omega_0^4 \omega_n^3 \mu_0 A F}{D_n^2 D_{2n}}, \tag{23a}$$

Similarly, for the difference frequency generation, the frequency $\omega_m$ is negative and, using explicitly a negative sign in Eq. 20, we obtain

$$\chi^{(2)}_{yyy}(\omega_n - \omega_m; \omega_n, -\omega_m) = ia \frac{\omega_0^4 (\omega_n - \omega_m) \omega_n \omega_m \mu_0 A F}{D_n D_m^* D_{n-m}}, \tag{23b}$$

where $n$ and $m$ can take any values between $\pm \Lambda$. The susceptibilities $\chi^{(2)}_{yyy}(-2\omega_n)$ and $\chi^{(2)}_{yyy}(-(\omega_n - \omega_m))$ are straightforwardly seen to be the conjugates of Eqs. 23a and 23b.

Note that, according to Eq. (20), the nonlinear susceptibility $\chi^{(2)}_{yyy}(0; \omega_n, -\omega_n)$ that is responsible for the effect of optical rectification vanishes, despite the fact that the response $q^{(2)}(0)$ is nonzero according to Eq. (13). This is an expectable result since, for the type of metacrystals we consider, the magnetic susceptibility depends on the AC response of the medium, while the response $q^{(2)}(0)$ does not vary in time. The zero-frequency second order response, in combination with the linear response or a higher-order response does contribute though to the higher-order susceptibilities. This contribution is addressed in the Appendix, where we provide a more intuitive derivation of the higher-order response for some combinatorial frequencies.

As can be seen from Eq. (21), many more combinatorial frequencies arise as the result of third order nonlinear processes. We provide, as an example, the explicit expressions for the third order susceptibilities $\chi^{(3)}_{yyyy}(3\omega_n)$ responsible for the third harmonic generation, and $\chi^{(3)}_{yyyy}(\omega_n)$, relating the nonlinear response at the fundamental frequency $\omega_n$ aka power dependent refractive index.

For the susceptibility at the third harmonic, one has $\omega_n = \omega_m = \omega_p$ in Eq. (23), which gives

$$\chi^{(3)}_{yyyy}(3\omega_n; \omega_n, \omega_n, \omega_n) = 3 \frac{\omega_0^6 \omega_n^4 \mu_0^2 A^2 F}{D_n^3 D_{3n}} \left[ b - \frac{2a^2 \omega_0^2}{3 D_{2n}} \right]. \tag{24a}$$

The response at the fundamental frequency $\omega_n$ can result both from the interaction of a field at $\omega_n$ with itself, leading to the effect of self-phase modulation, and from the interaction of two distinct fields, leading to a cross-phase modulation process [14, 27]. For the first case, we set $\omega_m = \omega_n$ and $\omega_p = -\omega_n$ in Eq. 21, resulting in

$$\chi^{(3)}_{yyyy}(\omega_n;\omega_n,\omega_n,-\omega_n) = \frac{\omega_0^6 \omega_n^4 \mu_0^2 A^2 F}{D_n^3 D_n^*}\left[\frac{4a^2\omega_0^2}{3D_0} + \frac{2a^2\omega_0^2}{3D_{2n}} - b\right], \tag{24b}$$

while for a similar response leading to cross-phase modulation, using $\omega_p = -\omega_m$ in Eq. 21, we obtain

$$\chi^{(3)}_{yyyy}(\omega_n;\omega_n,\omega_m,-\omega_m) = \frac{\omega_0^6 \omega_n^2 \omega_m^2 \mu_0^2 A^2 F}{D_n^2 D_m D_m^*}\left[\frac{2a^2\omega_0^2}{3D_{n+m}} + \frac{2a^2\omega_0^2}{3D_{n-m}} + \frac{2a^2\omega_0^2}{3D_0} - b\right]. \tag{24c}$$

Note that Eq. 24c approaches 24b as $\omega_m$ approaches $\omega_n$, as expected for the susceptibilities $\chi^{(3)}_{yyyy}(\omega_n;\omega_n,\omega_n,-\omega_n)$ and $\chi^{(3)}_{yyyy}(\omega_n;\omega_n,\omega_m,-\omega_m)$ [14].

Using Eqs. 24b and 24c in Eq. 18c and performing the summation to account for the possible number of permutations leading to each response, we can also see that, as expected, the nonlinear magnetization produced by two distinct fields is twice larger than the one produced by a single field:

$$M^{(3)}(\omega_n;\omega_n,\omega_n,-\omega_n) = 3\chi^{(3)}_{yyyy}(\omega_n;\omega_n,\omega_n,-\omega_n)|H|_n^2 H_n, \tag{25a}$$

$$M^{(3)}(\omega_n;\omega_n,\omega_m,-\omega_m) = 6\chi^{(3)}_{yyyy}(\omega_n;\omega_n,\omega_m,-\omega_m)|H_m|^2 H_n. \tag{25b}$$

The effective permeabilities characterizing the response at other combinatorial frequencies can be obtained from Eqs. 20 and 21 in a similar manner. We emphasize again that no new geometrical or linear circuit parameters are introduced into the higher order terms; aside from factors relating to the nonlinear component ($a$ and $b$ in this case), all unknown coefficients can be found by analyzing the linear response.

According to Eqs. 19, 20, and 21, we can also alternatively express $\chi^{(2)}(2\omega)$ and $\chi^{(3)}(\omega)$ as products of linear magnetic susceptibilities at frequencies $2\omega$ and $\omega$, allowing for a more explicit relation between the linear and nonlinear susceptibilities. For example,

$$\chi^{(2)}_{yyy}(2\omega_n) = -ia\frac{\chi^{(1)}(\omega_n)\chi^{(1)}(\omega_m)\chi^{(1)}(\omega_n+\omega_m)}{\omega_n\omega_m(\omega_n+\omega_m)\mu_0 A^3 C_0^2 N^2} \tag{26}$$

and

$$\chi^{(3)}_{yyyy}(\omega_s;\omega_n,\omega_m,-\omega_p) = -ia\frac{\chi^{(1)}(\omega_n)\chi^{(1)}(\omega_m)\chi^{(1)}(\omega_p)\chi^{(1)}(\omega_n+\omega_m+\omega_p)}{\omega_n\omega_m\omega_p(\omega_n+\omega_m+\omega_p)\mu_0 A^4 C_0^3 N^3}$$

$$\times\left[b - \frac{2a\chi^{(1)}(\omega_n+\omega_m)}{3A^2 N\mu_0 C_0(\omega_n+\omega_m)^2} - \frac{2a\chi^{(1)}(\omega_n+\omega_p)}{3A^2 N\mu_0 C_0(\omega_n+\omega_p)^2} - \frac{2a\chi^{(1)}(\omega_n+\omega_m)}{3A^2 N\mu_0 C_0(\omega_n+\omega_m)^2}\right]. \tag{27}$$

In particular, for the susceptibilities at the second harmonic and at the fundamental frequency,

$$\chi^{(2)}_{yyy}(2\omega_n) = \frac{-ia}{2N^2\omega_n^3 A^3\mu_0 C_0^2}\chi^{(1)}_{yy}(2\omega_n)\left[\chi^{(1)}_{yy}(\omega_n)\right]^2, \tag{28}$$

and

$$\chi^{(3)}_{yyyy}(\omega_n;\omega_n,\omega_n,-\omega_n) = \frac{1}{N^3\omega_n^4 A^4\mu_0 C_0^3}\left\{\frac{4a^2}{3} + \frac{\chi^{(1)}_y(2\omega_n)}{6N\omega_n^2 A^2\mu_0 C_0} - b\right\}\left[\chi^{(1)}_{yy}(\omega_n)\right]^3\chi^{(1)}_{yy}(-\omega_n) \tag{29}$$

### 2.3. Intensity-dependent permeability

Besides the effects of self- and cross-phase modulation, the third-order susceptibility at the fundamental frequency also leads to an intensity-dependent refractive index [14] (in our case, due to the intensity-dependent permeability), and, as discussed later, to a shift in the resonance frequency. The expression for the intensity-dependent permeability can be obtained by considering the total nonlinear magnetization at the fundamental frequency expanded up to the third order. Assuming a single incident field at $\omega_n$, we have according to Eqs. 17 and 18

$$M_y^{tot}(\omega_n) = \chi_{yy}^{(1)} H(\omega_n) + 3\chi_{yyyy}^{(3)} |H(\omega_n)|^2 H(\omega_n) \equiv \chi_{eff}(\omega_n) H(\omega_n), \tag{30}$$

so that the intensity-dependent permeability is given by

$$\mu_{eff}^{NL} \equiv 1 + \chi_{eff} = 1 + \frac{F\omega^2}{D_n} + 3\frac{F\omega_0^6 \omega^4 A^2 \mu_0^2 |H|^2}{D_n^3 D_{-n}} \left( \frac{4a^2 \omega_0^2}{3D_0} + \frac{2a^2 \omega_0^2}{3D_{2n}} - b \right). \tag{31}$$

We can expect from Eq. 31 a shift in the spectral position of the resonance with increased intensity compared to the linear case. Numerical examples of this shift will be presented in the following section, in which we apply the present theory to a particular case of a VLSRR-based medium.

## 3. Application to the VLSRR medium

A convenient means of achieving a nonlinear medium is to connect a varactor diode across the gap of an SRR. Depending on the number and orientation of the diodes integrated into the medium, different nonlinear properties will result. We investigate two specific configurations, one in which a single diode is inserted into the gap of an SRR, and the other in which two diodes are inserted into two gaps in the SRR, as shown in Figs. 2a and 2b. In the latter case, the varactors are oriented in a back-to-back configuration to prevent dissipative currents from flowing at large power levels, as occurs in the single-gap VLSRR [25].

1. *Single-gap VLSRR*

The nonlinear properties of individual VLSRRs have been analyzed theoretically and experimentally by Wang et al. [21], who demonstrated power-dependent resonance frequency shifting and bistable behavior. The nonlinearity associated with the diode elements occurs through the voltage dependent capacitance of the varactors. For the particular varactors studied in [21] and used here (Skyworks SMV1247), the junction capacitance has the form [39]

$$C(V_D) = C_0 \left(1 - V_D/V_p\right)^{-K}, \tag{32}$$

where $C_0$ is the zero bias capacitance, $K$ is the gradient coefficient and $V_P$ is the intrinsic potential. For small field amplitudes (and hence voltages), the time-dependent charge that accumulates as a function of the voltage can be found as

$$Q_D(V_D) = \int_0^{V_D} C(v)\,dv = \frac{C_0 V_p}{1-K}\left[1-\left(1-\frac{V_D}{V_p}\right)^{1-K}\right]. \tag{33}$$

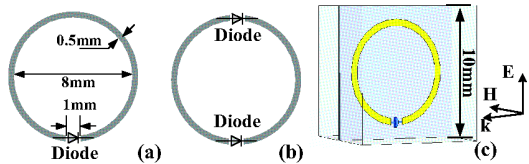

Fig. 2. Unit cell for the single-gap (a) and double-gap (b) metacrystals. (c) orientation of the unit cell with respect to the incident field.

Because we are primarily interested in the AC currents driven in the SRR, we make use of $C(V_D) = dQ_D/dV_D$, which relates the time-varying charge $Q_D$ to the time-varying voltage. Integrating Eq. (31) thus yields

$$V_D(q) = V_p\left[1-\left(1-q\frac{1-K}{V_p}\right)^{\frac{1}{1-K}}\right], \tag{34}$$

The diode thus provides a nonlinear charge/voltage relationship expressed as $V_D(q)$ that can be Taylor expanded for small arguments.

Assuming a small argument expansion in Eq. 32, we obtain the following expressions for the nonlinear coefficients that appear in Eq. 5:

$$a \equiv -\frac{\kappa}{2V_P},\quad b \equiv \frac{\kappa(2\kappa-1)}{6V_P^2}. \tag{35}$$

For the particular case of the Skyworks SMV1247 varactor, according to the specifications, $K=0.8$, $V_p = 1.5$ V [19], and hence $a = -0.2667\ s^{-1}V^{-1}$ and $b = 0.0356\ s^{-2}V^{-2}$.

If only one gap is present in the VLSRR, then the expressions derived in Section II for the linear and higher order susceptibilities can be used in their existing form, with the coefficients given by Eq. (35). The orientation of the unit cell relative to the incident field is shown in Figs. 2c.

2. *Double-gap VLSRR*

The single gap VLSRR is not optimal, since the current flowing in the SRR increases substantially with increasing power, leading to the voltage across the diode substantial enough to provide forward bias, hence increasing the resistive loss. To avoid the excitation of large currents, the diodes can be integrated into the SRR in an opposing configuration, such that one of the diodes is always reverse biased. The orientation of the back-to-back diodes is illustrated in Fig. 2b. Equation 1 in this case becomes

$$L\frac{dI}{dt} + R_s I + V_{D1} + V_{D2} = \varepsilon(t) \tag{36}$$

where $V_{D1}$ and $V_{D2}$ are the voltages at each of the effective capacitors.

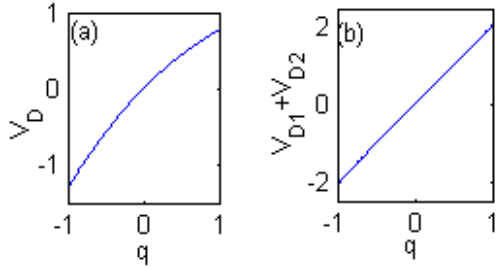

**Fig. 3 (a) The dependence of the voltage across a single varactor on the normalized charge; (b) The total voltage across both varactors as a function of charge for the double-gap case**

Due to the opposite configuration of the varactors, the same incident field will induce a forward bias on one of the varactors and the reverse bias on the other. Thus, the back-to-back configuration leads to the opposite sign of the induced voltage at the varactors and also to the opposite sign of the accumulated charge at the effective capacitors representing the varactors. Assuming both varactors are the same otherwise, this leads to the condition $V_{D_2}(q) = -V_{D_1}(-q)$ in Eq. (34). Accounting for this condition, Eq. (3) can be written in terms of $q$ as follows

$$\frac{d^2 q}{dt^2} + \gamma \frac{dq}{dt} + \omega_0^2 V_D(q) - \omega_0^2 V_D(-q) = \omega_0^2 \varepsilon(t). \tag{37}$$

As a result, the sum $V_{D_1} + V_{D_2}$ becomes an odd function of the normalized charge $q$, as illustrated in Fig 3b, and the even-order terms in the Taylor expansion of $V_{D_1} + V_{D_2}$ in Eq. 36 are canceled. Consequently, there is no contribution to $\chi_{yyyy}^{(3)}$ from the second-order susceptibility. Equation 5 in this case becomes

$$\ddot{q} + \gamma \dot{q} + \omega_d^2 q + \omega_d^2 b q^{(3)} = \omega_0^2 \varepsilon(t). \tag{38}$$

where $\omega_d \equiv \sqrt{2}\omega_0$ is the new resonant frequency of the circuit and $\omega_0$ is the resonant frequency from the single-gap case. Following the same procedure as before, we arrive at the following expression for the third-order susceptibility

$$\chi_{yyyy}^{(3)}(\omega_r; \omega_n, \omega_m, \omega_p) = +i2b \frac{\omega_0^8 (\omega_n + \omega_m + \omega_p) \omega_n \omega_m \omega_p \mu_0^3 A^4 N C_0}{D_d(\omega_n) D_d(\omega_m) D_d(\omega_p) D_d(\omega_s)}, \tag{39}$$

where $D_d(\omega) \equiv \omega_r^2 - \omega^2 - i\gamma\omega$. In particular, Eq. 24b for the third order susceptibility at the fundamental frequency becomes

$$\chi_{yyyy}^{(3)}(\omega_n; \omega_n, \omega_n, -\omega_n) = -2b \frac{F \omega_0^6 \omega^4 A^2 \mu_0^2}{D_d(\omega)^3 D_d(-\omega)}. \tag{40}$$

3. *Examples*

Figure 4 shows examples of the second and the third order nonlinear magnetic susceptibilities and the effective permeability obtained according to Eqs. 23a for the single-gap case and according to Eqs. 24a, 24b, and 26 for both single- and double-gap configurations, with the nonlinear coefficients $a$ and $b$ given by Eq. 35. In order to estimate the values of $\gamma$ and $F$ for the construction of Fig. 4, we numerically simulate the linear propagation of a plane wave through a single layer of the metacrystal with the following parameters: the unit cell size 10 mm, the diameter of the ring and the width of the metal strip, respectively, 8 mm and 0.5 mm, and the gap size 1 mm, as shown in Fig. 2. The substrate of thickness 0.2 mm is assumed to be made of FR4 with ε = 4.4(1+0.02i). The simulations are performed using the frequency domain solver contained in the commercial finite element package CST Microwave Studio.

From the computed transmission and reflection coefficients of a single layer of metacrystal, we employ a standard retrieval procedure [36] to find the frequency dependent constitutive parameters. We calculate $F$ and $\gamma$, as well as the resonant frequency, by fitting the resonance curve of the effective permeability retrieved via this procedure.

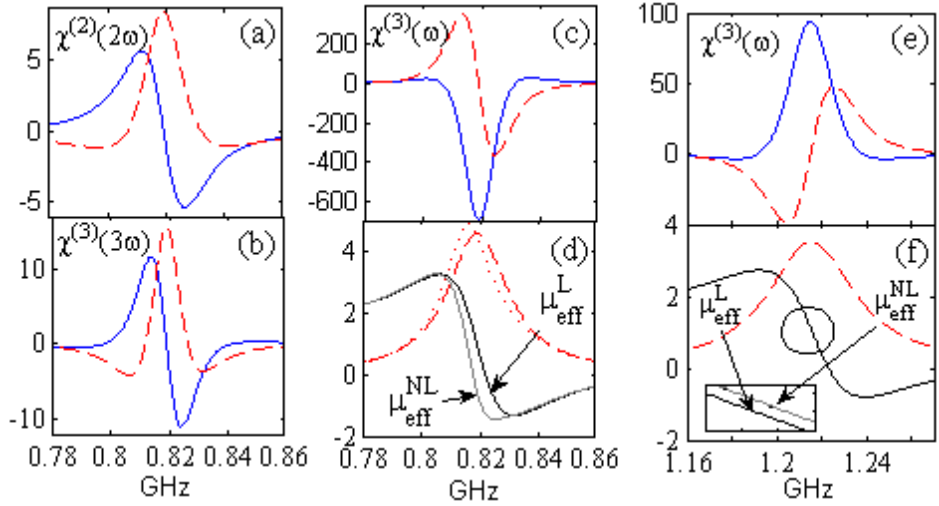

**Fig. 4.** Effective nonlinear susceptibilities and permeabilities for the single-slit (a-d) and the double-slit (e,f) metacrystals. The red dashed or dotted lines represent the imaginary part. The inset in (f) shows tha enlarged circled region on the graph. In calculating $\mu_{eff}^{NL}$, the incident field amplitude of $H(\omega_n) = 27$ mA/m and $H(\omega_n) = 21$ mA/m was used for the single- and double-gap cases, respectively.

Note that, although we do not take into account any interactions between the metamaterial elements in the above theory (i.e., no Lorentz factor), the simulations intrinsically take into account the entire periodic system (at least in the plane perpendicular to propagation). Thus, the numerical solution combined with the fitting procedure will correctly account for the effective medium properties of the metacrystal, including effects of spatial dispersion if such are present [40].

From the numerical simulations applied to the unit cell in Fig. 2, we obtain the values of $\gamma = 0.1202 \times 10^9$ $s^{-1}$, $F = 0.1426$ and $\omega_0 = 5.15 \times 10^9$ $rad/s$ for the single-gap metacrystal. Note that, assuming the area of the ring is given by the parameters in Fig. 2 ($A \approx 5 \times 10^{-5}$ $m^2$), the estimated value of $F$ leads to a slightly different zero bias capacitance value of $C_0 = 1.6$ pF than in the varactor specifications (2.4 pF [39]). This discrepancy is not unreasonable since the actual value of $C_0$ is modified by the packaging capacitance, the parasitic capacitance of the varactor, and the capacitance of the SRR gap itself. For the double-gap medium, the fitting parameters for the linear permeability are $F = 0.1362$, $\omega_r = 7.65 \times 10^9$ $rad/s$ and $\gamma = 0.2953 \times 10^9$ $s^{-1}$.

As seen from Fig. 4, the values of $\chi^{(2)}$ and $\chi^{(3)}$ near the resonance are many orders of magnitude larger than in standard nonlinear medium. Noting that the metacrystals are inherently resonant materials and hence the resonance region is important to consider, it is necessary to estimate the range of the field strength for which the truncation of the series expansion of the normalized voltage $q$ or the magnetization $M$ at the 3rd order of the field strength would be valid. We provide such as estimate in the following section. The maximum applicable field amplitude ensuring the convergence of the series expansion in each of the single-gap or the double-gap configurations was used in constructing Figs. 4.

As seen from Fig. 4e and from the inset in Fig. 4f, and as discussed earlier in connection with Eq. 31, the position of the effective permeability resonance shifts with the increased power. According to Eq. 31, the theory predicts the opposite direction of the resonant frequency shift for the two types of media considered, noting the opposite signs of $\chi^{(3)}(\omega)$ for the single- and double-slit metacrystals, as observed in Figs 2 b and d. The opposite directions of the resonance shift are observed in Figs. 4e and 4f and are in agreement with the directions obtained from the exact numerical solution of Eqs. 5 and 36 and from the experimental results, as discussed later in Sections IV and V.

Another observation from Fig. 4 is that the third order susceptibility for the third harmonic generation is an order of magnitude smaller than the $\chi^{(3)}(\omega)$, as seen in Figs. 4b and 4c. This indicates that we can expect a much effect of third harmonic generation that the third order response at the fundamental frequency, leading to the resonant frequency shift. A weaker response at the third harmonic clearly follows from Eqs. (24a) and (24b), since all the factors $D_n$ are resonant in the denominator in the susceptibility $\chi^{(3)}(\omega)$, while this is not the case for the third harmonic susceptibility $\chi^{(3)}(3\omega)$, in agreement with the discussion in Ref. [25] for the second-order response. The latter observation is general for the response at the fundamental frequency comparing to any other combinatorial frequency for a medium possessing a single magnetic resonance: while the nonlinear response of a metacrystal is resonant at any frequency combination, a generally larger response can always be expected at the fundamental resonance.

## 4. Limits of validity

The diode system we consider experimentally exhibits strong nonlinearity, and we a working close to the resonance of the effective medium, so that it is a legitimate concern as to whether the series expansions used in Eqs. 4-6 are convergent or appropriate. Specifically, since we seek to describe a medium by its second and third order nonlinear susceptibilities, it is important to determine if the higher order terms can be neglected. To address this concern, we perform a rough estimate by finding the value of the field strength that ensures the values of the fourth and the fifth terms in the series expansion are small relative to the second and the third order terms at the corresponding frequencies (the necessary convergence criteria). The estimated value of the field strength can then be verified by comparing the solution obtained by the perturbation approach with the exact numerical solution of Eq. 5. We proceed in this way, deriving the expressions for the fourth and the fifth orders of the magnetic susceptibility of the SRR medium in a similar way as above. The derivation and the expressions for the fourth order response at $2\omega$ and the fifth order response at $\omega$ are provided in the Appendix.

Comparing the contributions to the amplitude of the magnetization response at $\omega$ coming from the terms $3\chi^{(3)}(\omega)H^3$ and $10\chi^{(5)}(\omega)H^5$, we find that the maximum applicable field satisfying the necessary convergence criteria for the series truncated at the third order of the field strength is about 27 mA/m. The same field strength ensures the necessary convergence condition for the response at $2\omega$ truncated at the second order of the field strength. From an examination of the higher order terms in the series expansion for the double-gap VLSRR metacrystal, the maximum applicable $H$ for up to the third order expansion in field magnitude is estimated to be about 100 mA/m. The larger validity range is expected in this case since the higher-order effective permeability terms decrease more rapidly in the absence of the contributions from the even orders.

To investigate further the accuracy of the solution we solve numerically Eq. 3 using the exact expression for the voltage $V_D$ given by Eq. 32 and using the expansion of $V_D$ up to the third order in the power of $q$, according to

$$\frac{d^2q}{dt^2}+\gamma\frac{dq}{dt}+\begin{cases}\omega_0^2 V_D(q)\\ \omega_0^2 q + a\omega_0^2 q^2 + b\omega_0^2 q^3\end{cases} = \omega_0^2 \mathrm{E}_{dr}\cos(\omega t) \qquad (41)$$

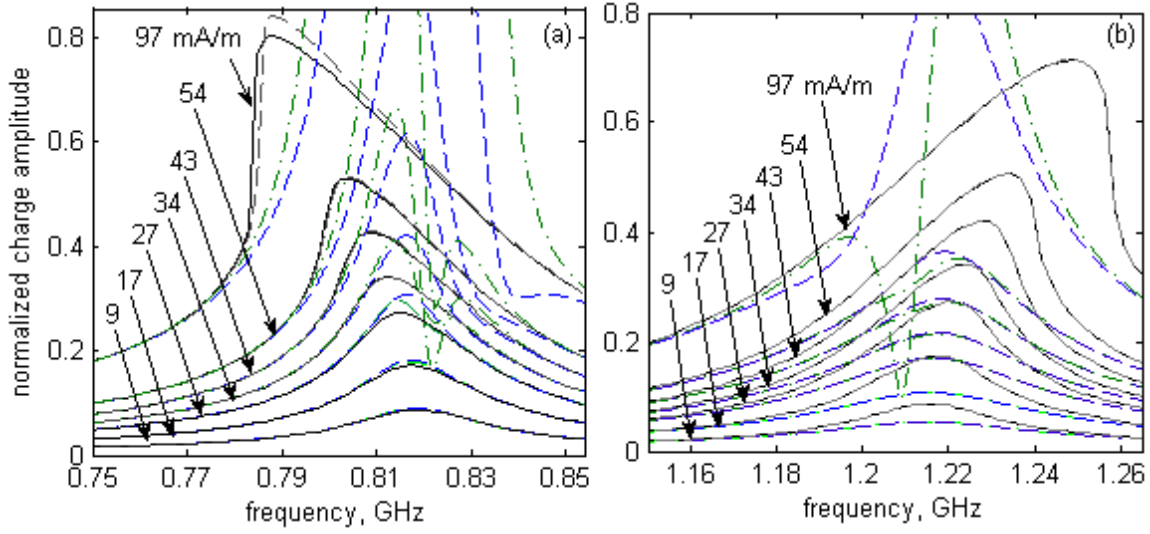

Fig. 5. Comparison of the numerical solution of the nonlinear oscillator equation employing the exact expression for VD (solid black) and $V_D$ expanded in Taylor series up to the 3rd order (dashed gray) with the solution obtained by the perturbation approach, assuming the expansion of the nonlinear magnetization up to the 3rd order (dashed blue) and the fifth order (dashed-dot green) in field strength. (a) Single-gap medium. (b) Double-gap medium.

and compare both results with the perturbative solution for the normalized voltage $q$. In Eq. 41, the driving voltage $E_{dr}$ is given by $E_{dr} \equiv AB_r\omega = 2\mu_0 H_0 A\omega$. We solved the differential equation 42 using the the ODE45 subroutine in Matlab (Mathworks), scanning the frequency $\omega$ of the applied field over the resonant region. For each $\omega$, we determine the amplitude of the response at the fundamental frequency $\omega$ and at each of the harmonics to obtain the resonant curve. We also solve an equation similar to Eq. 39 for the double-gap case, assuming $V_D(q)$ is the sum of voltages at the two capacitors, using $a = 0$ in its Taylor expansion, and replacing $\omega_0$ with $\omega_r$ in the linear and the cubic terms, as discussed above.

Figure 5 shows the comparison of the numerical and the analytical solutions for the fundamental frequency with several values of the incident power. For the single-gap case (Fig. 5a), the agreement is excellent for low magnetic field amplitudes up to 27 mA/m, in agreement with the value obtained earlier by analyzing the convergence of the series expansion. We also see good agreement for amplitudes up to about 2 dBm for the double-gap VLSRR metacrystal. For larger power values, the perturbative solution starts to deviate considerably from the numerical one near the resonance region, although it still exhibits very good agreement off-resonance. Accounting for the 5th order term in the perturbative solution, as indicated by the green dash-dot lines in Figs. 5, improves slightly the resulting amplitude and the position of the resonance, while leading to a stronger discrepancy in the resonance shape. The numerical solutions obtained using the exact expression for $V_D$ and using its expansion in power series just up to the third order, shown in grey dashed lines, agree very well with each other for all power values, indicating that the inaccuracy in analytical solution at high powers is due to the approximations involved with using the perturbation approach rather than with the Taylor expansion of the $V_D$ itself.

As discussed before and as seen both from Figs. 5a and 5b, the third order response at the fundamental frequency produces a shift in the resonant frequency of the metacrystal. The resonance shifts in opposite directions for the single-gap and the double-gap mediums, in agreement with the results predicted by the expression for the intensity-dependent permeability (Eq. 29) and discussed in the previous section. The opposite direction of the resonance shift obtained from the numerical solution for the two types of media can be directly predicted from Eq. 5 and 36, noting the opposite signs of the nonlinear coefficients in these equations (since $a<0$ and $b>0$, the resulting nonlinear term is negative, due to and $a\gg b$, in the single-gap case but is positive, due to $a=0$, for the double-gap case).

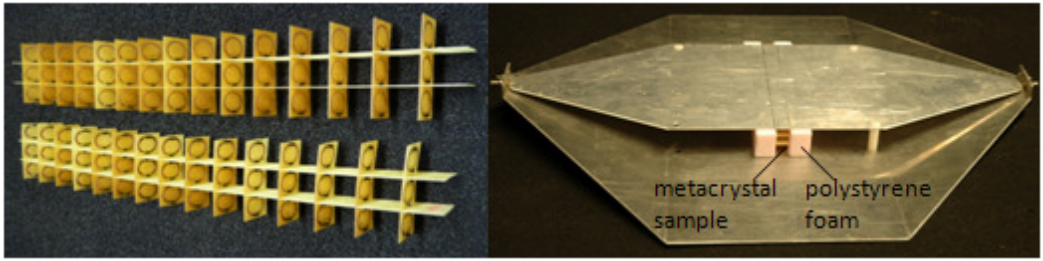

**Fig. 6. VLSRR medium and the transmission line with the metacrystal sample.**

The resonance in the solution for the normalized voltage $q$ and, consequently, in the effective nonlinear permeability characterizing the medium, would manifest itself as the minimum in the transmission spectrum of a plane wave propagating through the metacrystal. Therefore, the accuracy of the obtained solutions can be verified experimentally by observing the shift in the transmission minimum as the intensity of the plane wave propagating through the varactor-loaded metacrystal is increased and comparing the amount of the resonance shift with the one predicted by the theory and the numerical solution. We provide such a comparison with an experiment in the following section.

## 5. Experimental comparisons

To provide an experimental verification of the expressions derived above, a VLSRR medium was constructed, one unit cell in thickness (along the propagation direction) and 3x15 elements along the lateral directions. VLSRR metacrystals were constructed with the same unit cell parameters used in the above analysis, for both the single-and double-gap VLSRR cases. The metacrystal samples of the two kinds are shown in Fig 6a. A transmission line supporting TEM wave propagation below 2 GHz was used to measure the transmittance through the metacrystal samples, as shown in Figs. 6b. An Agilent vector network analyzer (PNA N5230A) was used to launch microwaves into the transmission line and detect the transmitted fields. The frequency-dependent transmission properties associated with both the waveguide and connecting cables were removed using a standard calibration method. The excitation power $P_{wg}$ from the network analyzer ranged from $-10$dBm to $+15$ dBm. The total loss from the transmission line structure, including the connection cables and adapters, was measured to be between 4 and 6 dB, so that the actual power exciting the sample ranged from $-14$dBm to 11 dBm, which spans the field range considered valid in the analytical calculations above. We note however that the actual power inside the waveguide and the corresponding magnetic field are accurate up to about 2 dBm due to the imperfect loss measurement and the approximate mode area used in the calculation.

The resulting transmission spectrum for the above range of powers is depicted in Fig. 7. We see that, in agreement with the theoretical predictions, the transmission minimum shifts with increasing power towards smaller frequencies for the single-gap metacrystal and towards larger frequencies for the double-gap medium. In the case of a single-gap metacrystal, we also see a decrease in the amplitude of the response with increased power. This is the consequence of the dissipative current running through the varactor [25], which is not accounted for in the present analytical model.

In Fig. 8, we present a comparison of the experimentally obtained resonant frequency dependence of the single- and double-gap VLSRR metacrystals on the incident power with the exact numerical solution of Eq. 39 and with the perturbative solution expanded up to the third and the fifth orders in field strength. The agreement between the exact numerical solution and the experimentally obtained resonant frequency is very good for the single-gap medium, validating the use of the effective circuit model leading to the nonlinear oscillator equation for the analytical description of the metacrystals. The slightly higher rate of the resonant frequency decrease with power seen in the experimental data in Fig. 8a compared to the numerical results can be attributed to the contribution from the nonlinear resistance to the resonant frequency shift as mentioned above. For the double-gap metacrystal, a slightly higher rate of the resonant frequency shift is observed in experiment than predicted by the numerical solution.

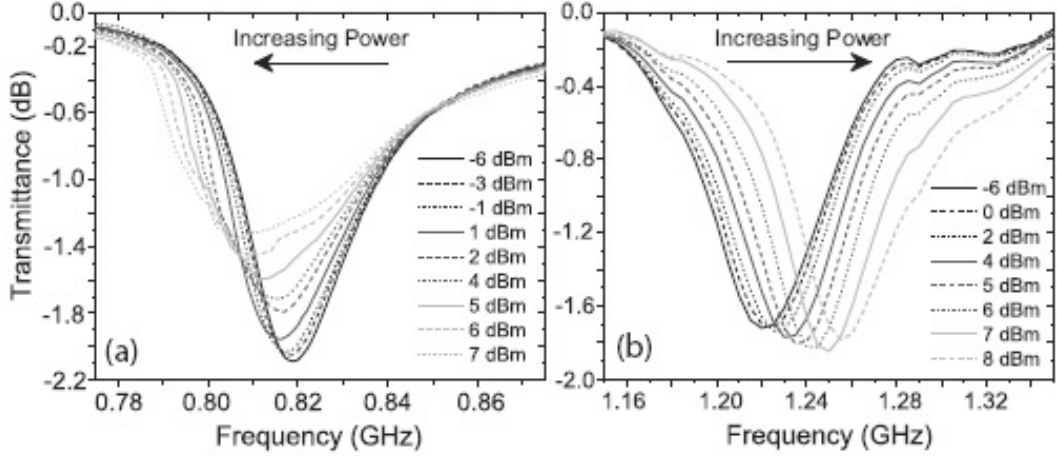

**Fig. 7.** Transmission coefficient for several incident power levels obtained experimentally for the (a) single-gap VLSRR medium and the (b) double-gap VLSRR medium.

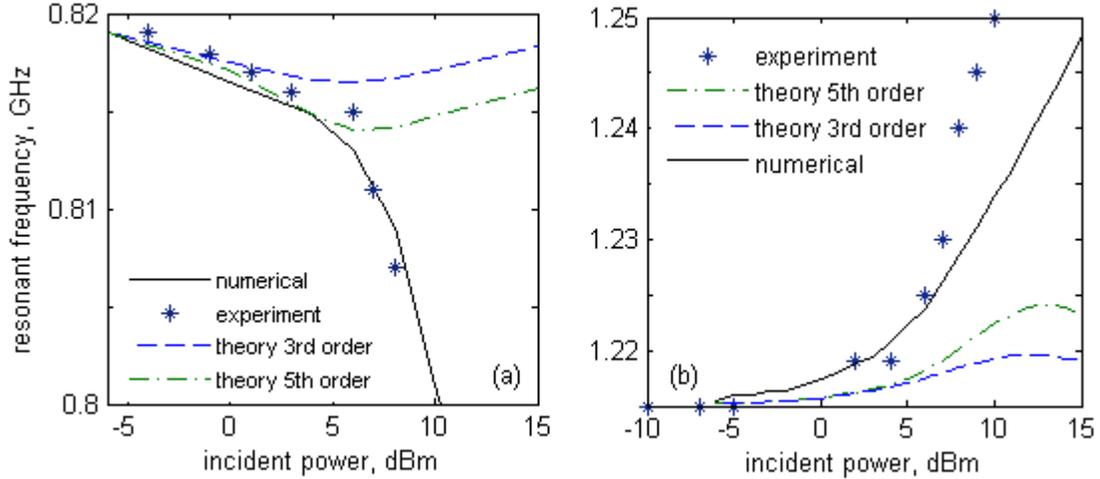

**Fig. 8.** Comparison of the nonlinear resonant frequencies vs. incident power obtained experimentally (stars), by the numerical solution of the nonlinear oscillator equation (solid black), and by the perturbation approach using the expansion up to the third (dashed blue) and the fifth (dashed-dot green) order of the field strength. (a) single-gap and (b) double-gap metacrystal.

The agreement with the solution obtained by the perturbative approach holds for the range of powers discussed in connection with Figs. 4 and 5, and accounting for the fifth order term in the series expansion leads to a small increase in the range of the validity of the approach, as seen from the green dashed lines in Fig. 8. According to Fig. 8b, the agreement between the experimental and the analytical resonant shift for the double-gap metacrystal holds up to about 2 dBm, corresponding to a field value of 21 mA/m, while the analytically predicted value is 100 mA/m. This difference can be attributed to the fact that the actual double-varactor unit cells are not ideally symmetric, leading to a non-ideal cancellation of the even-order terms in the voltage expansion. The contribution from these terms becomes pronounced once the field strength increases, producing a stronger resonance shift

One can verify that a similar discrepancy would hold near the resonance when comparing the perturbative and the exact solutions of a Drude-Lorentz oscillator at high power levels, using the expressions for the electrical nonlinear susceptibility in the case of natural materials at optical frequencies. For example, assuming the perturbative solution for a non-centrosymmetric material with the regular for optical case nonlinear parameters in the Lorentz model [15], the noticeable discrepancy near the resonance would start at intensity levels of about 1 TW/m$^2$. Normally, the Lorentz atom model is not the approach of choice for describing the resonant effects in natural materials at optical frequencies, and usually other techniques such as the two-level atom model [15] are used for the analysis of the resonant interaction of an optical field with a nonlinear medium. However, in case of the VLSRR metacrystals considered here, as seen from the comparison of the experimental transmission resonance shift and the exact numerical solution shown in Figs. 8, the nonlinear oscillator equation itself (Eqs. 5 and 36) provides a very close model describing the interaction of the electromagnetic wave with a metacrystal even at large power levels, both off- and near the resonance. Noting that the metacrystals considered here are inherently resonant materials, the modification to the perturbation approach allowing a better agreement with the numerical solution at high powers could be justified, but is beyond the scope of the present paper. For low power levels as discussed in this and the previous sections, the perturbative approach solution works well for characterizing nonlinear metacrystals in terms of second and third order susceptibilities.

## 6. Conclusions
We have employed a perturbative solution to the nonlinear oscillator model of the effective RLC-circuit of the unit cell to characterize the nonlinear properties of a metacrystal formed by the resonant elements that couple strongly to the magnetic field.. The nonlinear response of the metacrystals is characterized in terms of the series expansion of the nonlinear magnetization. We have provided general expressions for the effective magnetic nonlinear susceptibilities up to fifth order and discussed the valid power ranges for which the series expansion can be truncated at the third order of the field strength.

The expressions are convenient for the prediction, analysis, and possible enhancement of the nonlinear response of a metacrystal. In particular, while, according to the theory, for a medium exhibiting a single magnetic resonance the nonlinear response is resonant at all combinatorial frequencies, a generally stronger (an order of magnitude in the case of an VLSRR-based medium) nonlinear response is expected at the fundamental than at any other frequency, in agreement with the discussion in Ref. [25] for the case of the second-order susceptibility. The absence of optical rectification in the medium with a magnetic nonlinear response also follows directly from the presented theory.

Using a varactor-loaded metacrystal as an example, we analyzed in greater detail the third and the fifth order response at the fundamental resonant frequency, which leads to the effects of self-phase modulation, intensity dependent effective permeability, and resonant frequency shift. We compared the results for this response obtained by the perturbation solution with the results of the exact numerical solution of the nonlinear oscillator equation and with the experimental transmission results observing the shift of the resonant frequency of the metacrystal with the increased incident power. The comparison demonstrated that, near the resonance, the varactor loaded nonlinear metacrystals could be described by the effective second and third order nonlinear susceptibilities for the values of the excitation magnetic field amplitude up to 27 mA/m and 21 mA/m for the single-and double-gap metacrystals, respectively, corresponding to the incident powers of about 4 dBm and 2 dBm. There is also a very good off-resonance agreement between the perturbation approach and the numerical solutions for all power range considered (up to 10 dBm).

While the perturbative approach solution starts to deviate considerably near the resonance both from the numerical one and from the experimental results at higher powers, the agreement between the exact numerical solution of the same nonlinear oscillator equation and the experimental transmission results holds well even for high power values of up to about 10 dBm. This indicates that, in general, the nonlinear oscillator model provides a very close description of the interaction of the electromagnetic wave with a metacrystal at microwave frequencies even at high powers, both off- and near the resonance. Since commonly metacrystals are inherently resonant MM, the modification of the perturbation approach allowing a better agreement with the numerical solution at high powers is desired and is justified owing to the validness of the nonlinear oscillator model at high powers in general. In the range of the validness of the obtained solution, however, there is a very good agreement between the analytical, numerical, and experimental results for the metacrystal response, indicating that the analytical method can be potentially used to investigate other nonlinear phenomena in metacrystals.

**Acknowledgements**

This work was supported by the Air Force Office of Scientific Research (Contract No. FA9550-09-1-0562).

**Appendix A**

We derive the expression for the 4th and the 5$^{th}$ order effective permeabilities at, respectively, the second harmonic, $\chi^{(4)}(2\omega)$, and at the fundamental frequency, $\chi^{(5)}(\omega)$.

We assume up to the fifth order expansion of the voltage $V_D$ at the effective capacitor and of the normalized charge q in Eqs. (4) and (6) according to

$$V_D(q) = q + aq^2 + bq^3 + cq^4 + dq^5 \tag{A1}$$

$$\tilde{q}(t) = \lambda \tilde{q}^{(1)}(t) + \lambda^2 \tilde{q}^{(2)}(t) + \lambda^3 \tilde{q}^{(3)}(t) + \lambda^4 \tilde{q}^{(4)}(t) + \lambda^5 \tilde{q}^{(5)}(t) \tag{A2}$$

where, for the case of the VLSRR-based medium,

$$c \equiv -\frac{K(2K-1)(3K-2)}{24V_p^3}, \quad d \equiv \frac{K(2K-1)(3K-2)(4K-3)}{120V_p^4}, \tag{A3}$$

and the coefficients *a* and *b* are given by Eq. (33). Continuing the system of Eqs. (7), the expression for the forth and the fifth order response are

$$\ddot{\tilde{q}}^{(4)} + \gamma \dot{\tilde{q}}^{(4)} + \omega_0^2 \tilde{q}^{(4)} + \alpha \left[\tilde{q}^{(2)}\right]^2 + 2\alpha \tilde{q}^{(1)} \tilde{q}^{(3)} + 3\beta \left[\tilde{q}^{(1)}\right]^2 \tilde{q}^{(2)} + \upsilon \left[\tilde{q}^{(1)}\right]^4 = 0 \tag{A4}$$

$$\ddot{\tilde{q}}^{(5)} + \gamma \dot{\tilde{q}}^{(5)} + \omega_0^2 \tilde{q}^{(5)} + 2\alpha \tilde{q}^{(2)} \tilde{q}^{(3)} + 2\alpha \tilde{q}^{(1)} \tilde{q}^{(4)} + 3\beta \tilde{q}^{(1)} \left[\tilde{q}^{(2)}\right]^2 + 3\beta \left[\tilde{q}^{(1)}\right]^2 \tilde{q}^{(3)} + 4\upsilon \left[\tilde{q}^{(1)}\right]^3 \tilde{q}^{(2)} + \eta \left[\tilde{q}^{(1)}\right]^5 = 0, \tag{A5}$$

where $\upsilon$ and $\eta$ are related to the coefficients $c$ and $d$ as $\upsilon \equiv \omega_0^2 a$, $\eta \equiv \omega_0^2 b$. We assume for the purpose of this derivation a single plane-polarized incident wave at the frequency $\omega_n = \omega$. We consider now the last 4 terms in Eq. (A4) and account for all possible combination of frequencies in each of the terms contributing to the response at the second harmonic. For example, using the general solution for $\tilde{q}^{(2)}(t)$ given by Eq. (8b), and noting that $\omega_j$ in this equation can take values $2\omega$, 0, and $-2\omega$, the contribution of the $\alpha \left[\tilde{q}^{(2)}\right]^2$ term to the second order response in Eq. (A4) can come from the product $2a\omega_0^2 q^{(2)}(0) q^{(2)}(2\omega_j)$. Using similar considerations for the rest of the last 4 terms in Eq. (A4), the equation for the fourth order response at $2\omega$ can be written as

$$-4\omega_j^2 q^{(4)}(2\omega_j) - i2\omega_j \gamma \dot{q}^{(4)} + \omega_0^2 q^{(4)} + 2a\omega_0^2 q^{(2)}(0) q^{(2)}(2\omega) + 2a\omega_0^2 q^{(1)}(\omega) q^{(3)}(\omega) + 2a\omega_0^2 q^{(1)}(-\omega) q^{(3)}(3\omega)$$
$$+3\omega_0^2 b \left[q^{(1)}(\omega)\right]^2 q^{(2)}(0) + 6\omega_0^2 b q^{(1)}(\omega) q^{(1)}(-\omega) q^{(2)}(2\omega) + 4\omega_0^2 c \left[q^{(1)}(\omega)\right]^2 q^{(1)}(-\omega) = 0 \tag{A6}$$

We use Eqs. (10), (13), and (15) to obtain the expressions for the first, second, and third order response at each frequency contributing to the fourth order response in Eq. (A6). In particular, we have

$$q^{(2)}(2\omega) = a \frac{\omega_0^6 \omega^2 A^2 B^2}{D(2\omega) D^2(\omega)} \tag{A7}$$

$$q^{(2)}(0) = -2a \frac{\omega_0^6 \omega^2 A^2 B B^*}{D(0)D(\omega)D(-\omega)} \tag{A8}$$

$$q^{(3)}(\omega) = i \frac{\omega_0^8 \omega^3 A^3 B^2 B^*}{D(\omega)^3 D(-\omega)} \left( \frac{4a^2 \omega_0^2}{D(0)} + \frac{2a^2 \omega_0^2}{D(2\omega)} - 3b \right) \tag{A9}$$

$$q^{(3)}(3\omega) = -\frac{2a\omega_0^2}{D(3\omega)} q^{(1)}(\omega) q^{(2)}(2\omega) - \frac{b\omega_0^2}{D(3\omega)} \left[ q^{(1)}(\omega) \right]^3 \tag{A10}$$

where $D(\omega) \equiv \omega_0^2 - \omega^2 - i\gamma\omega_j$.

Using Eqs. (A7)- (A10) in Eq. (A6) to find $q^{(4)}(2\omega)$, and expressing $\chi^{(4)}(2\omega)$ according to $M^{(4)}(2\omega) = -iN2\omega q^{(4)}(2\omega) AC_0 = 4\chi^{(4)}_{yyyyy}(2\omega) H^3 H^*$, we obtain the following expression for the fourth order effective permeability contributing to the medium response at $2\omega$:

$$\chi^{(4)}_{yyyyy}(2\omega) = -i \frac{2NA^5 \mu^4 C_0 \omega^5 \omega_0^{10}}{D(2\omega_j) D^3(\omega) D(-\omega)} \left\{ 2c - 3\omega_0^2 ab \left[ \frac{1}{D(0)} + \frac{1}{D(2\omega)} + \frac{1}{3D(3\omega)} + \frac{1}{D(\omega)} \right] + \frac{2\omega_0^4 a^3}{D(2\omega)} \left[ \frac{1}{D(0)} + \frac{1}{D(3\omega)} + \frac{1}{D(\omega)} \right] + \frac{4\omega_0^4 a^3}{D(0) D(\omega)} \right\} \tag{A11}$$

Similarly as in Eq. (17) and (25), for the fifth order response at $\omega_n = \omega$, we have:

$$-\omega_j^2 q^{(5)}(\omega) - i\omega\gamma \dot{q}^{(5)} + \omega_0^2 q^{(5)} + 2a\omega_0^2 q^{(2)}(2\omega) q^{(3)}(-\omega) + 2a\omega_0^2 q^{(2)}(0) q^{(3)}(\omega) + 2a\omega_0^2 q^{(2)}(-2\omega) q^{(3)}(3\omega) + 2a\omega_0^2 q^{(1)}(\omega) q^{(4)}(0)$$
$$+ 2a\omega_0^2 q^{(1)}(-\omega) q^{(4)}(2\omega) + 3b\omega_0^2 q^{(1)}(\omega) \left[ q^{(2)}(0) \right]^2 + 6b\omega_0^2 q^{(1)}(-\omega) q^{(2)}(0) q^{(2)}(2\omega) + 6b\omega_0^2 q^{(1)}(\omega) q^{(2)}(-2\omega) q^{(2)}(2\omega)$$
$$+ 6b\omega_0^2 q^{(1)}(\omega) q^{(1)}(-\omega) q^{(3)}(\omega) + 3b\omega_0^2 \left[ q^{(1)}(\omega) \right]^2 q^{(3)}(-\omega) + 3b\omega_0^2 \left[ q^{(1)}(-\omega) \right]^2 q^{(3)}(3\omega) + 12c\omega_0^2 \left[ q^{(1)}(\omega) \right]^2 q^{(1)}(-\omega) q^{(2)}(0)$$
$$+ 12c\omega_0^2 \left[ q^{(1)}(-\omega) \right]^2 q^{(1)}(\omega) q^{(2)}(2\omega) + 4c\omega_0^2 \left[ q^{(1)}(\omega) \right]^3 q^{(2)}(-2\omega) + 10d\omega_0^2 \left[ q^{(1)}(\omega) \right]^3 \left[ q^{(1)}(-\omega) \right]^2 = 0 \tag{A12}$$

where $q^{(4)}(0)$ can be found from Eq. (A4) as follows

$$q^{(4)}(0) = -\frac{a\omega_0^2}{D(0)} \left\{ \left[ q^{(2)}(0) \right]^2 + 2q^{(2)}(2\omega) q^{(2)}(-2\omega) + 2q^{(1)}(-\omega) q^{(3)}(\omega) + 2q^{(1)}(\omega) q^{(3)}(-\omega) \right\}$$
$$- \frac{3b\omega_0^2}{D(0)} \left\{ 2q^{(1)}(\omega) q^{(1)}(-\omega) q^{(2)}(0) + \left[ q^{(1)}(\omega) \right]^2 q^{(2)}(-2\omega) + \left[ q^{(1)}(-\omega) \right]^2 q^{(2)}(2\omega) \right\} - \frac{6c\omega_0^2}{D(0)} \left[ q^{(1)}(\omega) q^{(1)}(-\omega) \right]^2. \tag{A13}$$

Expressing $q^{(5)}(\omega_j)$ from Eq.(A12) and using the correspondence $M^{(5)}(\omega) = -iN\omega q^{(5)}(\omega) AC_0 = 10\chi^{(5)}_{yyyyyy}(2\omega) H^3 H^{*2}$, we arrive at the following expression for the fifth order magnetic permeability contributing to the response at $\omega$ as follows:

$$\chi^{(5)}_{yyyyyy}(\omega_j) = \frac{A^6 C_0 N \mu^5 \omega^6 \omega_0^{12}}{D^4(\omega) D^2(-\omega)} \left\{ -d + 2a^2 c\omega_0^2 \left[ \frac{1}{5D(-2\omega)} + \frac{9}{5D(0)} + \frac{1}{D(2\omega)} \right] + \frac{9}{5} b^2 \omega_0^2 \left[ \frac{1}{2D(-\omega)} + \frac{1}{D(\omega)} + \frac{2}{D(0)} \right] \right.$$
$$-\frac{3}{5} a^2 b\omega_0^4 \left[ \frac{1}{D(0)} \left( \frac{4}{D(-\omega)} + \frac{8}{D(\omega)} + \frac{9}{D(0)} \right) + \frac{1}{D(-2\omega)} \left( \frac{1}{D(-\omega)} + \frac{1}{D(0)} + \frac{1}{3D(3\omega)} \right) \right.$$
$$\left. + \frac{1}{D(2\omega)} \left( \frac{1}{D(-\omega)} + \frac{8}{D(\omega)} + \frac{2}{D(2\omega)} + \frac{5}{D(0)} + \frac{1}{D(-2\omega)} + \frac{5}{3D(3\omega)} \right) \right]$$
$$+ \frac{4}{5} a^4 \omega_0^6 \left[ \frac{1}{D(0)} \left( \frac{2}{D(-\omega) D(0)} + \frac{1}{D(2\omega) D(-\omega)} + \frac{1}{D(-2\omega) D(-\omega)} + \frac{4}{D(0) D(\omega)} + \frac{4}{D(2\omega) D(\omega)} + \frac{1}{D^2(0)} + \frac{1}{D^2(2\omega)} + \frac{1}{2D(2\omega) D(-2\omega)} \right) \right.$$
$$\left. \left. + \frac{1}{D(2\omega)} \left( \frac{1}{2D(-2\omega) D(-\omega)} + \frac{1}{D(2\omega) D(\omega)} + \frac{1}{D(2\omega) D(3\omega)} + \frac{1}{2D(-2\omega) D(3\omega)} \right) \right] \right\}. \tag{A14}$$

We now use the parameters obtained in Section III for VLSRR medium to compare the 4th order response at $2\omega$ given by Eq. (A14) with the 2nd order response at the same frequency given by Eq. (A7) for a certain value of field strength to identify the value of field amplitude leading to the reduction of the forth order response compared to the second order.. We similarly compare the fifth and the third order responses at $\omega$. The value of the field amplitude leading to the reduction of both 4th and 5th order terms compared, respectively, to the 2nd and the 3rd order terms, is found to be about 21 mA/m.

The above derivation is done for the single-gap medium. The same expressions can be used for the double-gap case taking $b=0$ in Eqs (A4)-(A14) and using $D_d(\omega) \equiv \omega_0^2 - \omega_r^2 - i\gamma\omega_r$ in place of $D(\omega)$. In this way, the strength of the field providing a reduction of the 4th and the 5th order terms compared to the, respectively the 2nd and the 3rd order is about 100 mA/m.